\documentstyle[sprocl,epsf]{article}

\def\lsim{\mathrel{\rlap{\lower4pt\hbox{\hskip1pt$\sim$}}
    \raise1pt\hbox{$<$}}}         %less than or approx. symbol
\def\gsim{\mathrel{\rlap{\lower4pt\hbox{\hskip1pt$\sim$}}
    \raise1pt\hbox{$>$}}}         %greater than or approx. symbol

\def\beq{\begin{equation}}
\def\eeq{\end{equation}}
\def\bea{\begin{eqnarray}}
\def\eea{\end{eqnarray}}
\def\br{{\bf r}}
\def\brp{{\bf r^{\prime}}}

\def\lsim{\mathrel{\rlap{\lower4pt\hbox{\hskip1pt$\sim$}}
    \raise1pt\hbox{$<$}}}         %less than or approx. symbol
\def\gsim{\mathrel{\rlap{\lower4pt\hbox{\hskip1pt$\sim$}}
    \raise1pt\hbox{$>$}}}         %greater than or approx. symbol
\def\beq{\begin{equation}}
\def\eeq{\end{equation}}
\def\bea{\begin{eqnarray}}
\def\eea{\end{eqnarray}}
\begin{document}

\begin{flushright}
FZ-IKP(TH)-2000-11
\end{flushright}
\vspace{0.5cm}
\title{Probing the QCD pomeron in high-energy $\gamma^*\gamma^*$
 collisions
{\footnote {Talk at XXXVth Rencontres de Moriond, 
Les Arcs, Savoie, France, March 18-25, 2000}}}

%\author{A. B. AUTHOR, C. D. AUTHOR}

%\address{World Scientific Publishing Co, 1060 Main Street, 
%River Edge,\\ NJ 07661, USA\\E-mail: wspc@wspc.com} 

\author{V. R. ZOLLER}

\address{Institute for  Theoretical and Experimental Physics,\\
 Moscow 117218, Russia\\E-mail: zoller@heron.itep.ru}

\maketitle\abstracts{
 Based on the color dipole representation, we investigate 
consequences for the $\gamma^{*}\gamma^{*},\gamma^{*}\gamma$ scattering 
of the finding by Fadin, Kuraev and Lipatov that incorporation of 
asymptotic freedom into the BFKL equation makes the QCD pomeron a series 
of isolated poles in the angular momentum plane. We present parameter-free
predictions for the vacuum exchange contribution to the photon structure
function which agree well with OPAL and L3 determinations. A good agreement
is found between our  predictions for the energy and photon 
virtuality  dependence of the photon-photon cross section 
$\sigma^{\gamma^*\gamma^*}(W,Q^2,P^2)$ and the recent data taken by the 
L3 Collaboration.}
%........................................................
%.........................................................
%..........................................................
High-energy virtual photon-photon scattering can be viewed as 
interaction of small size color dipoles from the beam and target photons, 
which makes $\gamma^{*}\gamma^{*},
\gamma^{*}\gamma$ scattering at high energies (LEP, LEP200 \& NLC) an 
indispensable probe of short distance properties of the QCD pomeron 
exchange.

In this note we study  scattering of virtual and real photons
$\gamma^*(q) +\gamma^*(p) \to  X\,$
in the  regime of large
Regge parameter ${1\over x}$ 
\beq
{1 \over x}={W^2+Q^2+P^2\over {Q^2+P^2+m^2_{\rho}}}\gg 1,
\label{eq:1.2}
\eeq
where $W^2=(q+p)^2$ is the center-of-mass energy squared of colliding 
space-like photons $\gamma^*(q)$ and  $\gamma^*(p)$  with virtualities
$q^2=-Q^2$ and $p^2=-P^2$, respectively. The mass of the $\rho$ meson 
squared sets the natural scale for real photons.

In the color dipole (CD) basis the beam-target scattering is considered
 as interaction 
of color dipoles $\bf r$ and $\bf{r^{\prime}}$ in both  the beam ($b$) and 
target ($t$) particles.
 The fundamental 
quantity is  the forward dipole scattering amplitude and/or the 
dipole-dipole cross section $\sigma(x,\bf{r},\bf{r^{\prime}})$. Once 
$\sigma(x,\br,\brp)$ is known the total cross section of $bt$ scattering 
$\sigma^{bt}(x)$ is calculated as 
\beq
\sigma^{bt}(x)
=\int dz d^{2}{\bf{r}} dz^{\prime} d^{2}{\bf{r^{\prime}}}
|\Psi_b(z,{\bf{r}})|^{2} |\Psi_{t}(z^{\prime},{\bf{r^{\prime}}})|^{2}
\sigma(x,{\bf{r},\bf{r^{\prime}}}) \,.
\label{eq:2.1}
\eeq
In  (\ref{eq:2.1}) $\sigma(x,\bf{r},\bf{r^{\prime}})$ is beam-target symmetric
and universal for all beams and targets, the beam and target dependence is
concentrated in probabilities $|\Psi_b(z,{\bf{r}})|^{2}$ and
$|\Psi_{t}(z^{\prime},{\bf{r^{\prime}}})|^{2}$ to find a color dipole,
$\br$ and $\brp$ in the beam and target, respectively. Hereafter we focus
on cross sections averaged over polarizations of the beam and target photons.

 The incorporation of 
asymptotic freedom into the BFKL equation makes the QCD pomeron a series 
of isolated poles in the angular momentum plane\cite{FKL}.
 The contribution of each
isolated pole to the high-energy scattering amplitude satisfies the familiar
Regge factorization \cite{Gribov}, 
which in the CD basis implies the CD BFKL-Regge expansion
\beq
\sigma(x,r,r^{\prime})=\sum_{m}C_m\sigma_m(r)\sigma_m(r^{\prime})
\left({x_0\over x}\right)^{\Delta_m}\,.
\label{eq:2.3}
\eeq
Here the dipole cross section $\sigma_m(r)$ is an  eigen-function of the 
CD BFKL equation \cite{JETPLett,DER,PISMA1,NZZJETP,NZHERA,PION}
\beq
{\partial\sigma_{m}(x,r)\over \partial \log(1/x)}=
 {\cal K}\otimes \sigma_{m}(x,r)=\Delta_{m}\sigma_{m}(x,r),
\label{eq:2.4}
\eeq
with eigen value (intercept) $\Delta_{m}$.

Then, 
combining (\ref{eq:2.3}) and (\ref{eq:2.1}) and adding in the soft and
quasi-valence (reggeon) components, we obtain ($m={\rm soft},0,1,2,...$)
\beq
\sigma^{\gamma^*\gamma^*}(x,Q^2,P^2)=
{(4\pi^2 \alpha_{em})^{2} \over Q^{2}P^{2}}
\sum_{m} C_m f_m(Q^2)f_m(P^2)
\left({3x_0\over 2x}\right)^{\Delta_m}
+\sigma^{\gamma^*\gamma^* }_{\rm qval}\,.
\label{eq:2.5}
\eeq
For DIS off (quasireal) real photons, $P^{2}=0$,
\beq
F_{2\gamma}(x,Q^2)=\sum_m A^{\gamma}_m
f_m(Q^2)
\left({3\over 2}{x_0\over {x}}\right)^{\Delta_m}
+F_{2\gamma}^{\rm qval}(x,Q^2)\,.
\label{eq:F2.6}
\eeq
%%%%%%%%%%%%%%%%%%%%%%%%%%%%%%%%%%%%%%%%%%%%%%%%%%%%%%%%%%%%%%%
Here
$\sigma_m^{\gamma^*}(Q^2)=
 \langle {\gamma^*_T}  | \sigma_m(r)|{\gamma^*_T}
\rangle +  \langle {\gamma^*_L}  | \sigma_m(r)|{\gamma^*_L}
\rangle\,$
is calculated with the well known color dipole distributions in the 
transverse (T) and  longitudinal (L) photon of virtuality $Q^{2}$ derived 
in \cite{NZ91}, and the eigen structure functions are defined as
usual: 
$
f_m(Q^2)= {Q^2\over
 {4\pi^2\alpha_{\rm em}}}
 \sigma_m^{\gamma^*}(Q^2)\,.$
The analytical formulas for eigen structure functions $f_m(Q^2)$ are 
found in \cite{GammaGamma}.  We do not need any new parameters compared to 
those used
in the description of DIS and real photoabsorption on protons 
\cite{JETPLett,DER,PION}, the results
for the expansion parameters $A_{m}^{\gamma}$ and $\sigma_{m}^{\gamma}(0)$
are summarized in \cite{GammaGamma}.

\begin{figure} 
\epsfxsize=0.5\hsize
\epsfbox{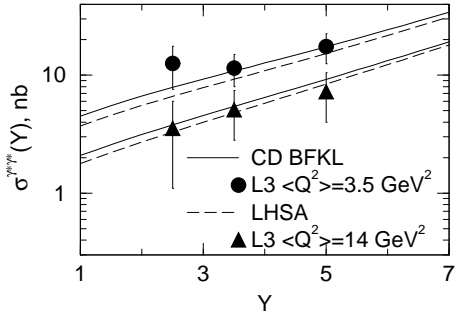}
\caption{Predicitions from CD BFKL-Regge expansion for the vacuum exchange
 component 
of the $\gamma^{*}\gamma^{*}$ cross section for the diagonal 
case of $\langle Q^{2} \rangle = \langle P^{2}\rangle $  are confronted
to the experimental data by the L3 Collaboration.} 
\label{fig1}
\end{figure}
 In fig.~1 we
compare our predictions to the L3 data \cite{L3SIGY} 
on the vacuum exchange contribution to $\gamma^{*}
\gamma^{*}$ scattering.  
The solid curve is a result of the complete BFKL-Regge expansion for the
vacuum exchange, the dashed curve is a sum of the rightmost hard BFKL
exchange and soft-pomeron exchange. The agreement of our estimates 
with the experiment is good, the contribution of subleading hard BFKL 
exchange is negligible within the experimental error bars.

The discussion of the photon structure function (SF) follows closely that
of the proton and pion SF's in \cite{JETPLett,DER,PION}.
There is a fundamental point that the distribution of small-size color 
dipoles in the photon is enhanced compared to that in the proton 
\cite{PION} which enhances the importance of the rightmost hard 
BFKL exchange.

Our predictions for the photon structure function are presented in 
fig.~2.
 At moderately small $x$ there is a substantial non-vacuum reggeon 
exchange contribution 
from DIS off quasi-valence quarks  which
can be regarded as well constrained by the large $x$ data, we use
here the GRS parameterization \cite{PHOTONGRS}. A comparison of the solid and 
dotted curves shows clearly that subleading hard BFKL exchanges are
numerically small in the experimentally interesting region of $Q^{2}$,
the rightmost hard BFKL pole exhausts the hard vacuum contribution 
for $2 \lsim  Q^{2} \lsim 100\, $ GeV$^{2}$.
\begin{figure}
\epsfxsize=0.7\hsize
\epsfbox{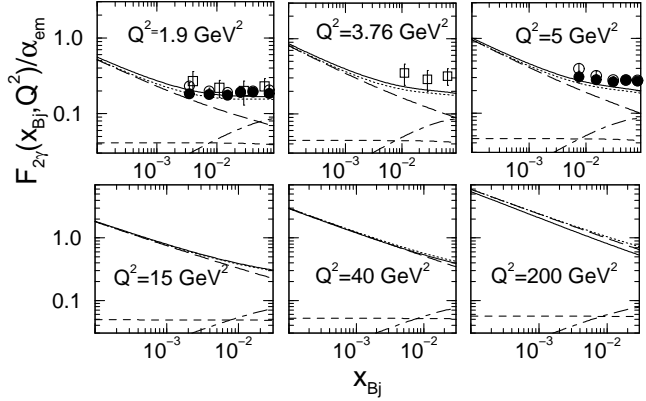}
\caption{Predicitions from CD BFKL-Regge expansion
  for the photon structure function.
 The solid curve shows the result from
the complete BFKL-Regge expansion the soft-pomeron (the dashed curve)
and valence (the dot-dashed curve) components included, the dotted 
curve shows the rightmost hard BFKL  (LH) plus soft-pomeron  (S) plus 
quasi-valence (V) approximation (LHSVA). The long dashed line corresponds to 
the LH plus S approximation (LHSA).  }
\label{fig2}
\end{figure}
The data on the photon structure function at sufficiently
 small-x \cite{L3,OPAL}.
 are in good agreement with the
predictions from the CD BFKL-Regge expansion. A comparison with the 
long-dashed curve which is the sum of the rightmost hard BFKL and soft
exchanges shows that the experimental data are in the region of 
$x$ and $Q^{2}$  still affected by non-vacuum reggeon (quasi-valence) exchange,
going to smaller $x$ and larger $Q^{2}$ would improve the sensitivity
to pure vacuum exchange greatly. \\

{\bf Acknowledgments: }
 This work was partly supported by the grants
INTAS-96-597 and INTAS-97-30494 and DFG 436RUS17/11/99.

%........................................................

\end{document}